\documentclass[aps,prl,twocolumn,superscriptaddress,showpacs]{revtex4}
\usepackage{graphicx}
\usepackage{graphics}
\usepackage{amsfonts}
\usepackage{amsmath}
\usepackage{epsfig}

\begin{document}
\catcode`\ä = \active \catcode`\ö = \active \catcode`\ü = \active
\catcode`\Ä = \active \catcode`\Ö = \active \catcode`\Ü = \active
\catcode`\ß = \active \catcode`\é = \active \catcode`\è = \active
\catcode`\ë = \active \catcode`\ô = \active \catcode`\ê = \active
\catcode`\ø = \active \catcode`\ò = \active \catcode`\í = \active
\catcode`\Ó = \active \catcode`\ú = \active \catcode`\á = \active
\catcode`\ã = \active
\defä{\"a} \defö{\"o} \defü{\"u} \defÄ{\"A} \defÖ{\"O} \defÜ{\"U} \defß{\ss} \defé{\'{e}}
\defè{\`{e}} \defë{\"{e}} \defô{\^{o}} \defê{\^{e}} \defø{\o} \defò{\`{o}} \defí{\'{i}}
\defÓ{\'{O}} \defú{\'{u}} \defá{\'{a}} \defã{\~{a}}



\newcommand{\li}{$^6$Li $ $}
\newcommand{\na}{$^{23}$Na $ $}
\newcommand{\cs}{$^{133}$Cs}
\newcommand{\kk}{$^{40}$K}
\newcommand{\rb}{$^{87}$Rb}
\newcommand{\vect}[1]{\mathbf #1}
\newcommand{\mf}{$m_F$}
\newcommand{\g}{g^{(2)}}
\newcommand{\one}{|1\rangle}
\newcommand{\two}{|2\rangle}
\newcommand{\limol}{$^6$Li$_2$}
\newcommand{\V}{V_{12}}
\newcommand{\kfa}{\frac{1}{k_F a}}

\title{Pairing without Superfluidity: The Ground State of an Imbalanced Fermi Mixture}

\author{C. H. Schunck}
\email{chs@mit.edu}
\affiliation{MIT-Harvard Center for Ultracold
Atoms, Research Laboratory of Electronics, Department of Physics,
Massachusetts Institute of Technology, Cambridge, Massachusetts
02139,USA}
\author{Y. Shin}
\affiliation{MIT-Harvard Center for Ultracold Atoms, Research
Laboratory of Electronics, Department of Physics, Massachusetts
Institute of Technology, Cambridge, Massachusetts 02139,USA}
\author{A. Schirotzek}
\affiliation{MIT-Harvard Center for Ultracold Atoms, Research
Laboratory of Electronics, Department of Physics, Massachusetts
Institute of Technology, Cambridge, Massachusetts 02139,USA}
\author{M. W. Zwierlein}
\affiliation{MIT-Harvard Center for Ultracold Atoms, Research
Laboratory of Electronics, Department of Physics, Massachusetts
Institute of Technology, Cambridge, Massachusetts 02139,USA}
\affiliation{present address: Institut f\"ur Physik, AG Quantum,
Staudinger Weg 7, 55128 Mainz, Germany}
\author{W. Ketterle}
\affiliation{MIT-Harvard Center for Ultracold Atoms, Research
Laboratory of Electronics, Department of Physics, Massachusetts
Institute of Technology, Cambridge, Massachusetts 02139,USA}

\date{\today}

\begin{abstract}
Radio-frequency spectroscopy is used to study pairing in the normal
and superfluid phases of a strongly interacting Fermi gas with
imbalanced spin populations. At high spin imbalances the system does
not become superfluid even at zero temperature. In this normal phase
full pairing of the minority atoms is observed. This demonstrates
that mismatched Fermi surfaces do not prevent pairing but can quench
the superfluid state, thus realizing a system of fermion pairs that
do not condense even at the lowest temperature.
\end{abstract}


\maketitle

Fermionic superfluidity has many manifestations in nature and occurs
in such diverse systems as superconducting materials, liquid $^3$He,
neutron stars, and ultracold quantum gases. At its heart lies the
formation of fermion pairs. While the Pauli principle forbids
identical fermions to occupy the same quantum state, pairs of
fermions can condense and thus become superfluid. Superconductivity,
the flow of electrical current without resistance, is a
manifestation of fermionic superfluidity in a condensed matter
system. Superconductors are characterized by a temperature $T^*$
where electrons start to pair and a critical temperature $T_c$ for
the onset of superconductivity. In conventional superconductors,
understood within the framework of Bardeen-Cooper-Schrieffer (BCS)
theory, fermion pairs form and condense simultaneously, i.e.
$T^*=T_c$. In high-temperature superconductors strongly correlated
electrons exist in the normal phase at $T^*>T_c$. The interactions
that mediate pairing and ultimately lead to superconductivity in
these complex systems are still under debate~\cite{lee06rev}.
Another strongly interacting, but comparatively simple fermion
system is an ultracold gas of neutral fermionic atoms. In these
gases, high-temperature superfluidity has been recently
observed~\cite{zwie05vortex}, opening a new approach to explore the
highly correlated normal phase of strongly interacting fermions and
its relation to the onset of superfluidity.

Ultracold atomic Fermi mixtures of two spin states close to a
Feshbach resonance realize a highly controllable model system for
strongly interacting fermions. By resonantly changing the
interaction strength between the fermionic atoms the crossover from
BCS superfluidity of loosely bound pairs to Bose-Einstein
condensation (BEC) of tightly bound molecules can be explored.
BEC-BCS crossover theory at finite temperature contains pairing in
the normal phase below a temperature
$T^*>T_c$~\cite{lee06rev,triv95,pera02pair,chen05rev}. Evidence for
pairing above $T_c$ in ultracold Fermi gases was found in
~\cite{chin04,kinn04gap} via radio-frequency (rf) spectroscopy. In
the present work, we use rf spectroscopy to study primarily the
normal state of an imbalanced spin mixture. An imbalance in the spin
populations of the two-state Fermi system leads to a qualitative
change of the phase diagram: above a certain, interaction dependent
population imbalance the transition to the superfluid state is
suppressed even at zero temperature. This is known as the
Chandrasekhar-Clogston (CC) or Pauli paramagnetic limit of
superfluidity~\cite{chan62,clog62}. In several works the CC limit is
assumed to imply pair dissociation and is referred to as ``Pauli
pair breaking''~\cite{cape02,mask02,kuma06fflo}, i.e. $T^*$ and
$T_c$ are assumed to vanish simultaneously.

The CC limit has been previously observed and characterized in
ultracold atomic gases~\cite{zwie06imb}. Here, we report on the
observation of a gap in a single-particle excitation spectrum
(representing a spin response function) of a highly imbalanced
sample. This implies that the system is in a correlated state and
that the minority component is almost completely paired. Pairing of
fermions is thus not necessarily a precursor to superfluidity: $T^*$
is finite even when $T_c$ vanishes. The CC limit of superfluidity,
at least for strong interactions, is not associated with breaking of
fermion pairs but only with the quenching of the superfluid state.
Another and probably very different system with finite $T^*$ and
vanishing $T_c$ has been discussed in strongly underdoped
cuprates~\cite{lee06rev}.

The rf spectra presented in this work were also correlated with an
indirect signature for superfluidity by determining pair condensate
fractions~\cite{rega04,zwie04pairs}. We conclude that rf spectra
cannot distinguish, at present experimental resolution, between
normal and superfluid states.

In the experiment a strongly interacting, imbalanced spin mixture of
$^6$Li fermions in the two lowest hyperfine states, labeled
$|1\rangle$ and $|2\rangle$ (corresponding to the
$|F=1/2,m_F=1/2\rangle$ and $|F=1/2,m_F=-1/2\rangle$ states at low
magnetic field) was created in an optical dipole trap at 833 G, the
center of the $|1\rangle - |2\rangle$ Feshbach resonance (see
refs.~\cite{zwie04pairs} and~\cite{hadz03big_fermi} for details). On
resonance all interactions in the $|1\rangle - |2\rangle$ mixture
are universal as the Fermi energy $E_F$ and the inverse Fermi
wavenumber $1/k_F$ are the only relevant energy and length scales.
The imbalance $\delta$ of the mixture was controlled as reported in
refs.~\cite{zwie06imb} and~\cite{shin06imb}, where
$\delta=(N_1-N_2)/(N_1+N_2)$ with $N_1$ and $N_2$ the atom number in
state $|1\rangle$ and $|2\rangle$, respectively. Here, $E_F$, $k_F$
and the Fermi temperature $T_F$ are given for a non-interacting
Fermi gas with the same atom number as the majority component. To
access a broader range of temperatures two optical traps with
different waists were used, characterized by the axial and radial
trapping frequencies $\omega_a$ and $\omega_r$ which are given in
the figure captions of the rf spectra.

\begin{figure}
\begin{center}
\includegraphics[width=2.8in]{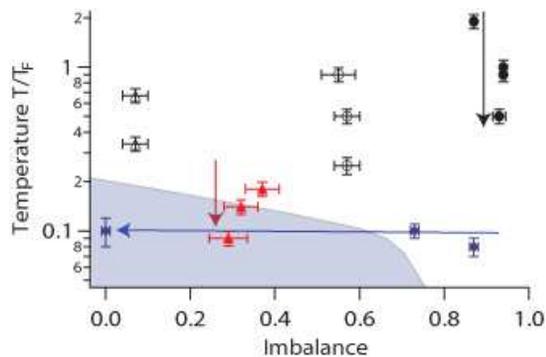}
\caption[Title]{The temperature-imbalance diagram shows where the rf
spectra presented in Fig.~2 (black circles), 4 \textbf{A$-$C} (blue
diamonds) and 4 \textbf{D$-$F} (red triangles) have been taken. All
spectra were obtained on resonance at 833 G. The arrows indicate the
order in which the spectra are displayed in the figures. As a guide
to the eye the shaded region indicates the superfluid phase. The
spectra corresponding to the open circles and triangles are similar
to the spectra of Fig.~2A to~2C and are shown in the supplemental
information. Except for the data close to zero imbalance, for which
the interacting temperature $T'$ is given, temperatures have been
determined from the non-interacting wings of the majority
cloud~\cite{zwie06direct}} \label{fig:fig1}
\end{center}
\end{figure}

\begin{figure}
\begin{center}
\includegraphics[width=3in]{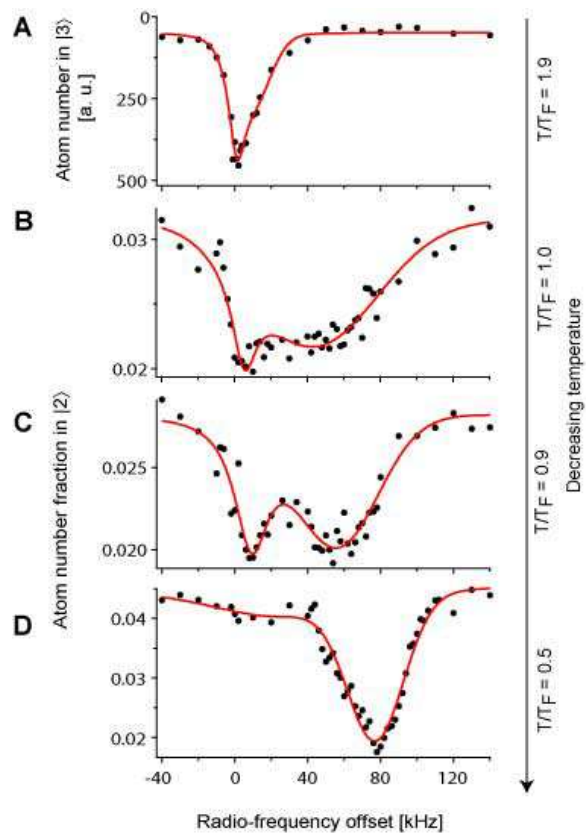}
\caption[Title]{RF spectroscopy of the minority component in an
imbalanced ($\delta\sim0.9$), strongly interacting mixture of
fermionic atoms above the Chandrasekhar-Clogston (CC) limit of
superfluidity. As the temperature is lowered full pairing develops
in the absence of superfluidity. \textbf{A}: An asymmetric and broad
peak centered at the position of the atomic line is observed. The
asymmetry and the large width might be caused by the presence of
pairing correlations already at $T/T_F=1.9$. Only for this spectrum
heating was applied and the atom number in state $|3\rangle$ was
recorded (see the supplemental information). \textbf{B} and
\textbf{C}: The pairing peak emerges. \textbf{D}: at $T/T_F=0.5$ the
pairing peak remains and the minority atoms are almost fully paired
(see also Fig.~4A). As a guide to the eye a Lorentzian fit to the
atomic line and a Gaussian fit to the pairing peak are included. The
spectra were taken for the following parameters (the black circles
in Fig.~1): A) $\delta=0.87$, $E_F=h \times 260$ kHz, $T/T_F=1.9$;
B) $\delta=0.94$, $E_F=h \times360$ kHz, $T/T_F=1.0$; C)
$\delta=0.94$, $E_F=h\times 360$ kHz, $T/T_F=0.9$; D) $\delta=0.93$,
$E_F=h\times 340$ kHz, $T/T_F = 0.5$, where $h$ is Planck's
constant. The trapping frequencies were $\omega_r = 2\pi\times 3.5$
kHz and $\omega_a = 2\pi\times 77$ Hz.} \label{fig:highimb}
\end{center}
\end{figure}

The interactions were spectroscopically probed in a three-level
system~\cite{gupt03rf}. A 2-ms rf pulse resonant with the transition
from state $|2\rangle$ (the minority component) to a third state,
labeled $|3\rangle$ ($|F=3/2, m_F=-3/2\rangle$ at low field) was
applied. Immediately after the rf pulse the optical trap was
switched off and the cloud was allowed to expand for absorption
imaging. Two absorption images of atoms in state $|2\rangle$ and
$|1\rangle$ were taken successively and the atom number fraction
$N_2/(N_1+N_2)$ was obtained as a function of the applied rf. The rf
spectra at the highest imbalances were taken with a population
transfer smaller than 3\% of the total number of atoms. The data
points in all spectra are the average of three independent
measurements. Temperature was adjusted by evaporation to different
depths of the optical trap followed by recompression. Spectra
presented as a data set were taken with the same final trap depth.
Fig.~1 provides an overview of the imbalances and temperatures at
which the rf spectra have been obtained. Specific details are given
in the figure captions and in the supplemental information. All
radio-frequencies were referenced to the $|2\rangle-|3\rangle$
resonance recorded in the absence of atoms in state $|1\rangle$.

The rf spectroscopy measures a single-particle spin excitation
spectrum for the minority component of the
mixture~\cite{kinn04rf,ohas05,he05rf,yu06delta}. To understand the
expected rf spectra one can use a simplified description of the gas
as a mixture of free atoms and molecule-like pairs which is strictly
valid only far on the BEC side of the Feshbach resonance.
Transferring an unbound atom from state $|2\rangle$ into state
$|3\rangle$ requires an energy $\Delta E_{23}$. As the
$|1\rangle-|3\rangle$ mixture is also strongly interacting due to a
$|1\rangle-|3\rangle$ Feshbach resonance located at 690
G~\cite{gupt03rf}, we first assume, as in refs.~\cite{chin04}
and~\cite{kinn04gap}, that mean field shifts (i.e. shifts
corresponding to Hartree terms) are absent in the rf spectrum. Then
$\Delta E_{23}$ and the width of the atomic $|2\rangle-|3\rangle$
transition are independent of the density of atoms in state
$|1\rangle$. If, however, an atom in state $|2\rangle$ is paired
with an atom in state $|1\rangle$, the rf photon has to provide the
binding energy $E_B$ required to break the pair in addition to
$\Delta E_{23}$. Therefore, if pairing is present in the system, a
second peak emerges in the minority rf spectrum that is separated
from the atomic line and associated with
pairing~\cite{chin04,kinn04gap}. In a Fermi cloud, pairing is strong
only near the Fermi surface. Since the rf photons can excite atoms
in the whole Fermi sea the observed spectral gap $\Delta\nu$ may
have to be interpreted as a pair binding energy averaged over the
Fermi sea. Indeed in the BCS limit one has $h \Delta\nu \propto
\Delta^2/E_F$, where $\Delta$ is the BCS pairing
gap~\cite{yu06delta}. Under these working assumptions we interpret
the emergence of a gap in the spectrum as a pairing effect.

The presence of pairing in the normal phase has been observed in the
rf spectra for a highly imbalanced mixture, with $\delta\sim 0.9$,
on resonance at 833 G (Fig.~2). At high temperature only the atomic
peak was present, and as the temperature was lowered, a second peak,
the pairing peak emerged and separated from the atomic peak. At
sufficiently low temperatures essentially only the pairing peak
remained. This behavior is qualitatively similar to what has been
observed in an equal mixture~\cite{chin04}. The spectral gap
$\Delta\nu$, i.e. the shift of the pairing peak relative to the
atomic line increases as the temperature is lowered. At the lowest
temperature of 0.08 $T/T_F$ (Fig.~4A) we measured a shift of 0.38
$E_F$.

All the spectra in Fig.~2 have been obtained at high imbalances
above the CC limit of superfluidity. Here the system cannot undergo
a phase transition to the superfluid state even at zero temperature.
For a trapped gas on resonance the CC limit is reached at a critical
imbalance of
$\delta_{\bf{c,exp}}=0.74(5)$~\cite{zwie06imb,shin06imb} in
agreement with a calculated value of
$\delta_{\bf{c,theory}}=0.77$~\cite{lobo06imb}. Strong pairing
without superfluidity occurred also on the BCS-side of the Feshbach
resonance (Fig.~3). Here the imbalance $\delta=0.88$ was high above
the critical imbalance of $\delta_{\bf{c,exp}}=0.6(1)$, as
previously measured around this interaction
strength~\cite{zwie06imb}.

\begin{figure}
\begin{center}
\includegraphics[width=2.5in]{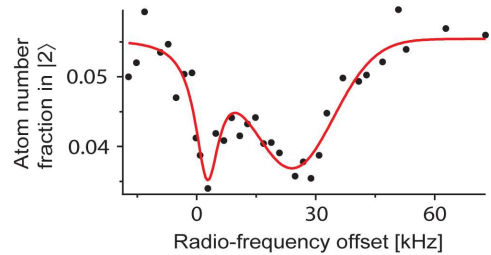}
\caption[Title]{RF spectrum of the minority component obtained at a
magnetic field of 937 G ($1/k_Fa_{12} = -0.18$) and imbalance
$\delta =0.88$, demonstrating strong pairing above the CC limit also
on the BCS side of the Feshbach resonance ($a_{12}$ is the $s$-wave
scattering length in the $|1\rangle - |2\rangle$ mixture). The rf
spectrum was taken for the following parameters: $E_F=h \times 280$
kHz and $T/T_F = 0.3$. The trapping frequencies were $\omega_r =
2\pi\times 2.9$ kHz and $\omega_a = 2\pi\times 64$ Hz.}
\label{fig:bcshighimb}
\end{center}
\end{figure}

As we have observed full pairing in the normal phase of the strongly
interacting gas, one might not expect the rf spectra to reveal the
onset of superfluidity. We have recorded rf spectra covering the
phase transition from the normal to the superfluid state by varying
imbalance (Fig.~4A$-$C) as well as temperature (Fig.~4D$-$F). In
both cases no signature of the phase transition is resolved,
although both the emergence of fermion pair condensates and sudden
changes in the density profiles~\cite{zwie06imb,shin06imb} show the
phase transition. In our previous work~\cite{zwie05vortex,zwie06imb}
these indirect indicators of superfluidity have been correlated with
the presence of quantized vortices, i.e. superfluid flow.

Figure~4A$-$C illustrates that working with high imbalances has the
advantage of reducing line broadening effects that arise from
averaging over the inhomogeneous density distribution of the sample.
The narrowest line was observed at the highest imbalance (Fig.~4A),
where the minority is considerably smaller than the majority cloud.
The homogeneous linewidth should reflect the wavefunction of a
single fermion pair. The observed narrow linewidth indicates
localization in momentum space well below the Fermi momentum $k_F$,
and hence a pairsize on the order of the interparticle spacing.

\begin{figure*}
\begin{center}
\includegraphics[width=6in]{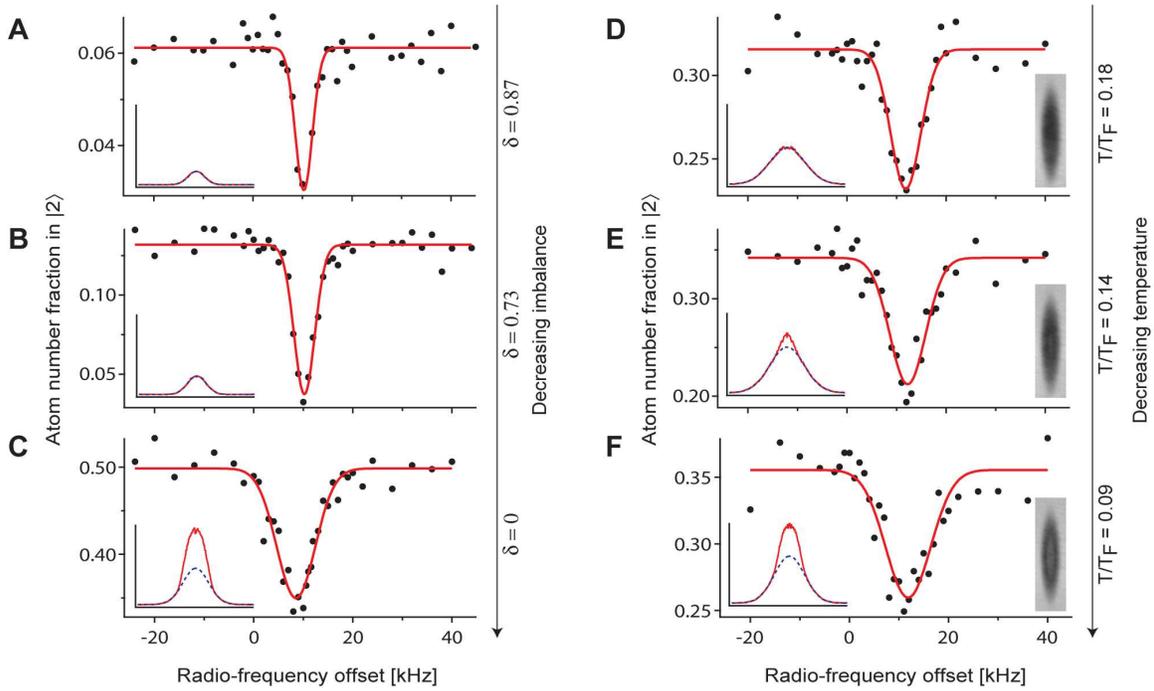}
\caption[Title]{RF spectra of the minority component obtained while
crossing the phase transition by reducing imbalance
(\textbf{A}$-$\textbf{C}) and temperature (\textbf{D}$-$\textbf{F}).
The rf spectra do not reveal the phase transition. The onset of
superfluidity is indirectly observed by fermion pair condensation.
The condensate fractions for \textbf{A} and \textbf{B} are zero and
35(2)\% in \textbf{C}. The onset of superfluidity as a function of
temperature occurs between \textbf{D} and \textbf{F}, with
condensate fractions of 0\% in \textbf{D}, 3(2)\% in \textbf{E}, and
17(3)\% in \textbf{F}. The left insets show the column density
profile (red) of the minority cloud after a rapid magnetic field
ramp to the BEC side and further expansion (see the supplemental
information). The blue dashed line is a Gaussian fit to the thermal
background. The right insets in \textbf{D}$-$\textbf{F} show phase
contrast images for a trapped cloud, obtained at imbalances of the
opposite sign. The spectra were taken for the following parameters
\textbf{A$-$C}: A) $\delta=0.87$, $E_F=h \times27$ kHz,
$T/T_F=0.08$; B) $\delta=0.73$, $E_F=h \times27$ kHz, $T/T_F=0.10$;
C) $\delta=0.00$, $E_F=h \times23$ kHz, $T'/T_F=0.10$. See also the
blue diamonds in Fig.~1. The trapping frequencies were $\omega_r =
2\pi\times 143$ Hz and $\omega_a = 2\pi\times 23$ Hz. For the
spectrum in \textbf{C} we quote the temperature $T'$ obtained from a
fit to the interacting Fermi gas (see the supplemental information);
\textbf{D$-$F}: D) $\delta=0.37$, $E_F=h \times38$ kHz,
$T/T_F=0.18$; E) $\delta=0.32$, $E_F=h \times38$ kHz, $T/T_F=0.14$;
F) $\delta=0.29$, $E_F=h \times35$ kHz, $T/T_F=0.09$. See also the
filled red triangles in Fig.~1. The trapping frequencies were
$\omega_r = 2\pi\times 192$ Hz and $\omega_a = 2\pi\times 23$ Hz.}
\label{fig:imbscan}
\end{center}
\end{figure*}

We now examine the assumptions underlying our interpretation of the
peaks in the rf spectra. In particular we address the question
whether our observations can distinguish between pairing
correlations and mean field effects. Indeed, mean-field-like shifts
are observed, for example in the rf spectrum of Fig.~2C where the
atomic line shows a shift of 0.03 $E_F$ to higher energy. Although
the $|1\rangle-|3\rangle$ interactions are in the unitary regime for
a typical value of $k_Fa_{13} \simeq -3.3$ (varying for example from
-3 to -3.6 across the minority cloud in Fig. 2C), they may not have
fully converged to their value at unitarity and thus cause the
observed shifts. Here $a_{13}$ is the $s$-wave scattering length in
the $|1\rangle - |3\rangle$ mixture. However, all shifts of the
atomic line are small compared to the size of the spectral gap of up
to 0.38 $E_F$ and are only seen in the presence of the pairing peak.
Fig.~7 displays all observed shifts of atomic and pairing peaks
versus temperature. While the shifts of the atomic line are small at
all temperatures, the shifts associated with the pairing peak start
rising below $T/T_F\sim1$, accompanied by a decrease in the weight
of the atomic line. In the intermediate temperature range where the
rf spectra show a double-peak structure, the pairing peak should
originate primarily from the higher density region in the center of
the cloud and the atomic peak from the low density wings. Therefore,
if one would normalize the data according to the local density of
majority atoms, the data points for the atom peaks would shift up in
$T/T_F$ by a factor between 1.5 and 5, the smaller factor reflecting
the cases of large imbalance, where the minority cloud is
considerably smaller than the majority cloud. As a result, near
$T/T_{F\tt{local}} = 0.5$, we have observed both atomic peaks and
pairing peaks, which is an indication for the local coexistence of
unpaired and paired minority atoms. However, in this possible
coexistence region, either the peak separation is small or one peak
has very small weight. Therefore more work is needed to study the
possibility of coexistence. An alternative interpretation assumes
single local peaks and a sudden onset of peak shifts below
$T/T_F\sim1$. Also the second scenario appears to be incompatible
with a local mean-field approximation: the mean field in the
unitarity limit should saturate when T approaches $T_F$ and not vary
strongly for $T < T_F$, since the relative momentum of two particles
in this regime is dominated by the Fermi momentum and not by the
thermal momentum. Furthermore a sudden onset of interactions would
likely affect the density distribution of the minority atoms.
However, the minority clouds observed in expansion are well fit by a
single Thomas-Fermi profile~\cite{zwie06direct}.

The BEC-side picture of a mixture of single atoms and molecules
seems to extend into the resonance region in the sense that fermion
pairs form high above the superfluid transition temperature and
possibly coexist locally with unpaired atoms. However, the fermion
pairs on resonance behave differently compared to ``real''
molecules: their binding energy increases with lower temperature and
higher atomic density. Most importantly fermion pairs above the CC
limit do not condense at low temperature as bosonic molecules would
do at any imbalance. While some extensions of BCS mean-field
theories to the imbalanced case do not predict pairing at imbalances
$\delta$ above the CC limit~\cite{chie06phase}, a survival of Cooper
pairs ``far from the transition region'' has been
predicted~\cite{alei97} for a superconducting system that is driven
into the normal, paramagnetic phase by Zeeman splitting.

The observed spectral gaps appear to be insensitive to the density
of the minority atoms (see Fig.~4A-C). At very high imbalances one
should indeed approach the limit of one minority atom immersed in a
fully polarized Fermi sea. In
refs.~\cite{lobo06imb,chev06univ,bulg06asy} the ground state energy
for this scenario has been calculated to be about -0.6 $E_F$, for
example by using a modified Cooper-pair wavefunction
ansatz~\cite{chev06univ}. These calculations do not provide an
excitation spectrum and do not distinguish between pairing
(correlation) energies and mean field (Hartree) terms. Therefore the
theoretical result cannot be directly compared to our spectroscopic
measurement of $h \Delta\nu = -0.38 E_F$ at $T/T_F=0.08$.

There is still a debate, whether superfluidity can occur for large
imbalances and low atom numbers in highly elongated geometries
~\cite{part06def}. In light of our findings, it may be important to
clearly distinguish between the effects of pairing and of
superfluidity. It has also been suggested that the presence of an
atomic peak next to the pairing peak in the minority cloud at zero
temperature and high imbalance could provide evidence for exotic
forms of superfluidity like the Fulde-Ferrel-Larkin-Ovchinnikov
state~\cite{kinn06imb}. However, for the parameters studied here,
the atomic peak is seen to disappear as the temperature is reduced
(see Fig.~2,~4A).

In conclusion, working with imbalanced Fermi gases, we were able to
study and characterize pairing in a situation where no superfluidity
occurs even at zero temperature. The spectral gap $\Delta\nu$
appears to be only weakly dependent on the imbalance. This suggests
that near unitarity certain pairing correlations in the superfluid
state are similar to those in a dilute cloud of minority atoms
immersed into the Fermi sea of the majority. Moreover, this implies
that the energetics which drives the normal-to-superfluid phase
transition is not simply the observed pairing energy. Further
studies of the strongly correlated normal state might yield new
insights into the microscopic physics of the superfluid state.

We thank Wilhelm Zwerger, Patrick Lee, Kathy Levin, and Qijin Chen
for stimulating discussions. We also thank Daniel Miller for a
critical reading of the manuscript. This work was supported by NSF
and ONR.

\section{Supplemental Information}
\subsection{Experimental Details}

\emph{Determination of the atomic reference line:} For the data
taken at the center of the $|1\rangle - |2\rangle$ Feshbach
resonance the resonance frequency of the $|2\rangle - |3\rangle$
transition in the absence of atoms in state $|1\rangle$ was
determined to be 81.700 MHz $\pm$ 1 kHz, corresponding to a magnetic
field of about 833 G. The FWHM of a Lorentzian fit to the resonance
peak was less than 1 kHz. These values reflect day to day
fluctuations and correspond to a magnetic field stability better
than 0.2 G. The resonance frequency of the $|2\rangle - |3\rangle$
transition on the BCS-side of the Feshbach resonance (Fig.~3) was
81.187 MHz $\pm$ 1 kHz (corresponding to a magnetic field of 936.5
G), determined in the absence of atoms in state $|1\rangle$.

\begin{figure}
\begin{center}
\includegraphics[width=2.5in]{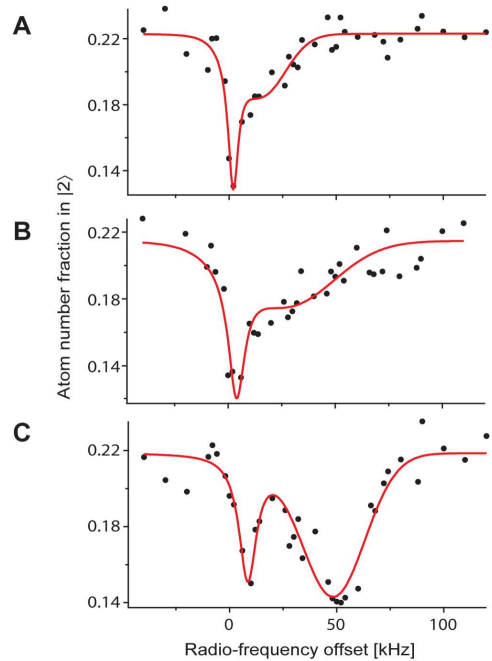}
\caption[Title]{Rf spectra of the minority component on resonance at
833 G. The spectra correspond to the open triangles shown in Fig.~1
of the letter and were obtained for the following parameters: A)
$\delta = 0.55$, $E_F = h \times 230$ kHz, $T/T_F = 0.9$; The
trapping frequencies for \textbf{A} were $\omega_r = 2\pi\times 3.4$
kHz and $\omega_a = 2\pi\times 76$ Hz. B) $\delta = 0.57$, $E_F = h
\times 230$ kHz, $T/T_F = 0.5$; C) $\delta = 0.57$, $E_F = h \times
220$ kHz, $T/T_F = 0.25$. The trapping frequencies for \textbf{B}
and \textbf{C} were $\omega_r = 2\pi\times 2.9$ kHz and $\omega_a =
2\pi\times 64$ Hz.} \label{fig:imb60}
\end{center}
\end{figure}

\emph{Rf pulse}:For all data a rf pulse of 2 ms was applied. This
pulse duration is optimized in terms of precision and minimizing a
dynamic response of the system during the rf pulse. For each
spectrum the rf power was adjusted to give an adequate
signal-to-noise ratio.

\emph{Determination of the atom number fraction in state
$|2\rangle$}: To obtain the atom number fraction $N_2/(N_1+N_2)$ two
absorption images, one of the minority component in state
$|2\rangle$ and the other of the majority component in state
$|1\rangle$, were taken successively. The time-of-flight before the
first absorption image as well as the delay time between the
absorption images were adjusted depending on the imbalance $\delta$
of the mixture, final temperature and the trapping frequency of the
optical dipole trap. The time-of-flight before the absorption image
of the minority varied between 200 $\mu$s and 8 ms, the delay time
between the images was in the range of 500 $\mu$s and 2 ms.

\emph{Imaging atoms transferred to state $|3\rangle$; Fig.~2A}: For
the rf spectrum in Fig.~2A, $T/T_F$ was increased by shortly
switching off the optical dipole trap and allowing for subsequent
equilibration before the rf pulse. The number of atoms transferred
to state $|3\rangle$ was recorded for a better signal-to-noise
ratio. The absorption image had to be taken within 200 $\mu$s after
applying the rf pulse. After longer time-of-flight atoms in state
$|3\rangle$ decayed through collisions. This precluded imaging atoms
in state $|3\rangle$ at lower temperatures where longer
time-of-flights were required before absorption imaging to avoid
saturation.

\emph{Weight of the atomic peak as function of imbalance}: The
population imbalance affects the weight of the atomic peak in rf
spectra obtained at the same $T/T_F$ (compare Fig.~2D,~5C and~6B).
As the imbalance decreases, the weight of the atomic peak increases.
This is likely due to the higher relative temperature compared to
the local binding energy in the the lower density region of the
majority could. That effect will result in a higher fraction of
unpaired atoms at small imbalances.

\emph{Temperature determination}: Except for equal and nearly equal
mixtures ($\delta<20\%$), temperatures were determined from the
\textit{non-interacting} wings of the majority cloud after
expansion~\cite{zwie06direct}. In ref.~\cite{zwie06direct} it was
found that for imbalances $\delta>20\%$ the non-interacting wings of
the majority cloud expand ballistically and are not affected by the
hydrodynamic expansion of the interacting component. For equal or
nearly equal mixtures the temperature $T'$ was determined directly
from a finite-temperature Thomas-Fermi fit to the whole density
profile of the interacting majority cloud. If on applies the
calibration of~\cite{kina05heat} the temperatures of (0.1,0.34,0.67)
$T'/T_F)$ in (Fig.~4C, Fig.~6B, Fig.~6A) should correspond to about
(0.1,0.23,0.45)$T/T_F$.

\emph{Chandrasekhar-Clogston limit}: The experimental value quoted
of $\delta_{\bf{c,exp}}=0.74(5)$ on resonance was obtained with the
following probes for superfluidity:  vortices and condensate
fractions~\cite{zwie06imb}, bimodal density distributions of the
minority cloud in time-of-flight~\cite{zwie06direct}. We would like
to emphasize, that the previous experimental determination of the
critical imbalance included a measurement of its \textit{temperature
dependence}, which was found to be weak at low
temperatures~\cite{zwie06imb}.

\emph{Condensate fractions}: Condensate fractions were obtained as
previously described
 in ref.~\cite{zwie04pairs} and \cite{zwie06imb}. The
samples were prepared as in the rf experiment, but the rf pulse was
not applied. Instead the gas was released from the trap and the
magnetic field was switched in 200 $\mu$s to 690 G, where the cloud
expanded for several ms. Then the magnetic field was ramped in 1 ms
to 720 G for absorption imaging. Condensate fractions were
determined from bimodal fits to the minority component. Condensates
were only observed when condensate fractions are explicitly stated
(Fig.~4).

\begin{figure}
\begin{center}
\includegraphics[width=2.3in]{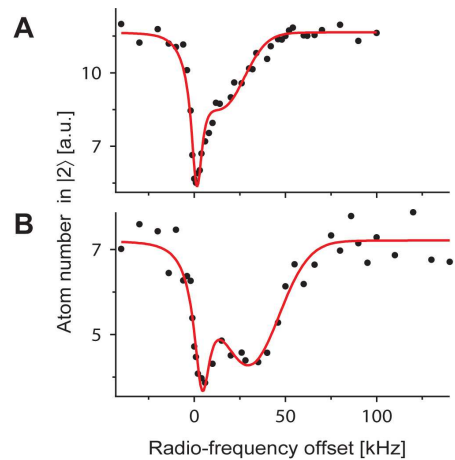}
\caption[Title]{Rf spectra of the minority component on resonance at
833 G. Since the majority component of the nearly equal mixture also
suffered significant losses after the rf pulse (probably due to
inelastic collisions), we report here the un-normalized atom number
in state $|2\rangle$ as a function of the applied radio frequency.
The spectra correspond to the open circles shown in Fig.~1 of the
letter and were obtained for the following parameters: A) $\delta =
0.07$, $E_F = h \times 210$ kHz, $T'/T_F = 0.67$; B) $\delta =
0.07$, $E_F = h \times 180$ kHz, $T'/T_F = 0.34$. The trapping
frequencies were $\omega_r = 2\pi\times 2.9$ kHz and $\omega_a =
2\pi\times 64$ Hz.} \label{fig:imb0}
\end{center}
\end{figure}

\begin{figure}
\begin{center}
\includegraphics[width=2.3in]{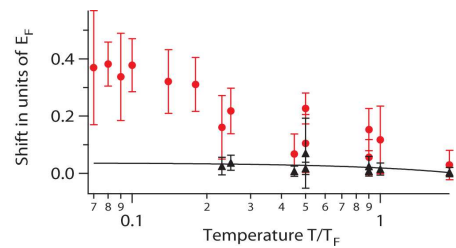}
\caption[Title]{Normalized shifts of the atomic peaks (black
triangles) and pairing peaks (red circles) as a function of $T/T_F$.
$E_F$ ($T_F$) is the Fermi energy (temperature) of a non-interacting
Fermi gas with the same number of atoms as the majority component.
The black line is a linear fit to the atomic peak shifts. The
temperatures $T'/T_F$ for equal or nearly equal mixtures were scaled
to $T/T_F$ (see temperature calibration). The error bars reflect the
full width at half maximum of a Gaussian fit to the peaks.}
\label{fig:shifts}
\end{center}
\end{figure}


\end{document}